# Optimization of the Design of the Hard X-ray Polarimeter X-Calibur

Qingzhen Guo[1,2*], Matthias Beilicke[1], Alfred Garson[1], Fabian Kislat[1], David Fleming[3], Henric Krawczynski[1]

[1] *Washington University in St. Louis, Department of Physics and McDonnell Center for the Space Sciences, 1 Brookings Dr., CB 1105, St. Louis, MO 63130, United States*
[2] *Northwestern Polytechnical University, State Key Laboratory of Solidification Processing, School of Materials Science and Engineering, Xi'an 710072, PR China*
[3] *University of Notre Dame, Department of Physics, Notre Dame, IN 46556, United States*

**Abstract**

We report on the optimization of the hard X-ray polarimeter X-Calibur for a high-altitude balloon-flight in the focal plane of the InFOCμS X-ray telescope from Fort Sumner (NM) in Fall 2013. X-Calibur combines a low-Z scintillator slab to Compton-scatter photons with a high-Z Cadmium Zinc Telluride (CZT) detector assembly to photo-absorb the scattered photons. The detector makes use of the fact that polarized photons Compton scatter preferentially perpendicular to the electric field orientation. X-Calibur achieves a high detection efficiency of order unity and reaches a sensitivity close to the best theoretically possible. In this paper, we discuss the optimization of the design of the instrument based on Monte Carlo simulations of polarized and unpolarized X-ray beams and of the most important background components. We calculate the sensitivity of the polarimeter for the upcoming balloon flight from Fort Sumner and for additional longer balloon flights with higher throughput mirrors. We conclude by emphasizing that Compton polarimeters on satellite borne missions can be used down to energies of a few keV.

## 1 Introduction

### 1.1 Scientific Motivation

X-rays are uniquely suited to study compact objects like neutron stars, pulsars, binary black hole systems, and soft gamma-ray repeaters as these objects are very bright in the X-ray energy band [1]. Furthermore, X-rays are well suited to study the continuum emission from the highly relativistic outflows from gamma-ray bursts and Active Galactic Nuclei. X-ray polarization measurement based on the photoelectric effect, the Compton effect, or pair production are one of the most exciting frontiers in contemporary astrophysics as they offer new diagnostics to test models explaining the X-ray emission from these sources. An X-ray polarimeter like X-Calibur measures not only the energies and arrival times of the X-ray photons, but obtains additional information about the energy-dependent polarization degree and polarization direction. Hard X-ray polarimetry can make substantial contributions to solving a number of important astrophysical questions [1], for example:

- Combining soft and hard X-ray polarimetry, the inclination angle of binary black hole systems, the orientation of the

*Corresponding author. Email address: qguo@physics.wustl.edu, Tel. 314 935 4745, Fax. 314 935 6219 (Q. Guo)

inner accretion disks, and the masses and spins of the black holes can be measured or constrained [2,3]. Contemporaneous observations over the broadest possible energy range are required to derive the best possible constraints. In general, the observations can also be used to test General Relativity in the strong gravity regime, although time resolved high signal-to-noise observations may be needed to break model degeneracies [4].

- Combined soft and hard X-ray observations can be used to study the accretion disks and the accretion disk coronae of Active Galactic Nuclei [3].
- Hard X-ray polarization measurements can give unique information about the particle acceleration processes in magnetars. The polarimetric observations can be used to study plasma effects and to experimentally probe quantum electrodynamic predictions that cannot be tested in terrestrial laboratories.
- Spectropolarimetric observations constrain the geometry and locale of the particle acceleration in pulsars and pulsar wind nebulae. As high-energy electrons lose their energy more rapidly than low-energy electrons, hard X-rays are produced in more compact regions than soft X-rays. As a consequence, higher polarization degrees are expected which constrain models effectively.
- Spectropolarimetric X-rays observations can probe the magnetic structure of the relativistic jets from Gamma-Ray Bursts (GRBs) and Active Galactic Nuclei. The observations have the potential to establish the helical structure of the jet magnetic field at the base of the jets, to constrain the uniformity of the magnetic field in the particle acceleration regions, and to distinguish between different emission mechanisms [5].
- X-ray polarimetry can be used to place stringent upper limits on the helicity dependence of the vacuum speed of light, and can thus be used for extremely sensitive tests of Lorentz Invariance [1,6].

For general reviews on the science drivers and techniques of X-ray polarimetry, the interested reader can consult [1,7,8,9] and references therein.

## 1.2 X-Calibur

X-Calibur is a polarimeter for the focal plane of an imaging mirror assembly similar to the ones used in the HERO [10], HEFT [11], InFOCμS [12] and NuStar [11,13] experiments. X-Calibur uses low-Z and high-Z materials to scatter and absorb incident X-ray photons, respectively. The design (Section. 3) aims at achieving the best possible polarization sensitivity in the hard X-ray (20 -75 keV) band. Three properties make it more sensitive than competing hard X-ray polarimeters:

- High Detection Efficiency: Most hard X-ray Compton polarimeters (e.g. POGO[14], GRAPE[15,16], and HXPOL[17]) use only 10~20% of the photons impinging on the detector assembly; some photons may hit absorbing rather than scattering detector elements, escape after scattering, or scatter too often to allow for a unique event reconstruction. By contrast, the detector configuration of X-Calibur can achieve >80% detection efficiencies over most of its energy range and all events can unambiguously be reconstructed.. This paper aims at optimizing the design of the X-Calibur polarimeter to achieve a performance limited by the physics of Compton scatterings and by the irreducible background alone.
- Low Background: X-Calibur achieves substantially lower background levels than other hard X-ray polarimeters that do not use focusing optics and rely on massive detector elements to collect the photons. Note that grazing incidence mirrors cause a specular reflection of the X-rays and change their polarization properties by less than 1% [18,19].

- Minimization and Control of Systematic Errors: X-Calibur will spin during the observations around the optical axis. The spinning makes it possible to distinguish φ-modulations caused by detector effects from φ-modulations caused by the polarization of the observed source. The azimuthal angle φ is the angle between the electric field vector of the incoming photon and the scattering plane. The analysis of the data in detector coordinates can be used to study the systematic effects, and the analysis in celestial coordinates to extract the physics results.

The X-Calibur design trades a high detection efficiency for imaging capabilities. This is well justified given that the most interesting targets for X-Calibur (accreting black holes and neutron stars) are too small to be spatially resolved by current technology. Below, we report on the sensitivity of X-Calibur when flown with different mirror assemblies on balloon-borne missions. As of writing this paper, NASA approved a one-day balloon flight of X-Calibur in the focal plane of the InFOCµS X-ray telescope [12] from Fort Sumner (NM, 34.47°N 104.24°W) for Fall 2013. The InFOCµS telescope will be equipped with a 40 cm diameter 8 m focal length Wolter type Al mirror assembly developed at Nagoya University. The mirror was flown already on the 2001 and 2004 InFOCµS balloon flights and comprises 255 double reflection shells, each made of 0.17 mm thick Al-foils with a multilayer coating for a broad energy range. The effective area of the mirror assembly is shown in the right panel of Fig. 4.

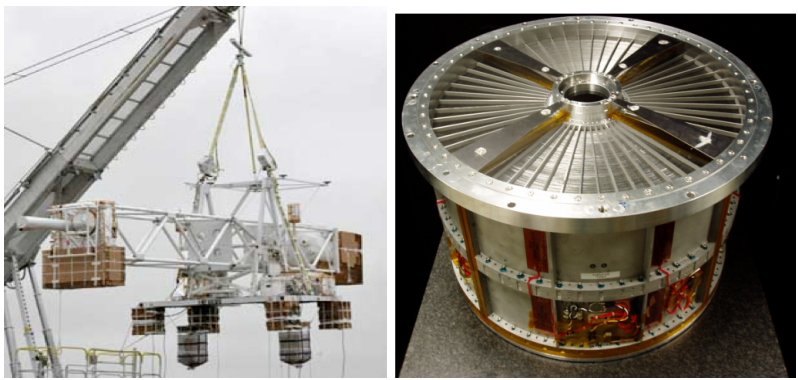

Figure: 1 Left panel: InFOCµS balloon gondola. The X-Calibur polairmeter will be located in the 8 m focal plane of the Wolter X-ray mirror (40 cm diameter). Right panel: the InFOCµS Al-mirror assembly. [12]

We envision longer follow-up duration balloon flights from the northern and southern hemispheres. An increased mirror area would lead to an increased signal rate while leaving the background rate almost unchanged – resulting in an improved signal-to-noise ratio. An attractive option is a longer duration balloon flight with a larger X-ray mirror assembly and makes it possible to achieve simultaneous broadband X-ray spectropolarimetric coverage in the 0.5 - 80 keV band.

In this paper, we present results from Monte Carlo studies of the polarimetric performance of X-Calibur and the optimization of the design based on the simulations. In Section 2, we introduce several general concepts important for X-ray polarimeters. We describe the design of X-Calibur in Section 3 and the simulations and the analysis tools in Section 4. Section 5 presents the results of the simulation study. Section 6 summarizes our conclusions and discusses the potential of a Compton polarimeter when used on a satellite borne mission.

## 2  General Considerations

In general terms, a Compton polarimeter like X-Calibur uses the fact that X-rays preferentially Raleigh, Thomson, or Compton scatter perpendicular to the orientation of their electric field vectors. If one accumulates many events from a linear polarized beam and determines the azimuthal scattering angle, the distribution will reveal a sinusoidal modulation with a 180° periodicity with an amplitude and phase depending on the polarization degree and direction, respectively. An important parameter describing a polarimeter is the amplitude of the azimuthal scattering angle distribution for a 100% linearly polarized signal. This amplitude is known as the modulation factor, µ, and is defined as:

$$\mu = \frac{C_{max} - C_{min}}{C_{max} + C_{min}} , \quad (1)$$

where $C_{max}$, $C_{min}$ refer to the maximum and minimum numbers of counts in the azimuthal scattering angle histogram. The performance of a polarimeter can be characterized by the Minimum Detectable Polarization (MDP) [8,1]. The MDP is the minimum degree of linear polarization which can be detected with a statistical confidence of 99% in a given observation time T:

$$MDP = \frac{4.29}{\mu R_{src}} \sqrt{\frac{R_{src} + R_{bg}}{T}} , \quad (2)$$

Here, $R_{src}$ and $R_{bg}$ are the source and background rates of events entering the analysis, and T is the integration time. We calculate error bars on the polarization degrees and directions of simulated energy spectra (see Fig. 15) based on the posteriori probabilities of the measured parameters (see [20,21] for related discussions).

## 3  The X-Calibur Design

X-Calibur uses a low-Z scattering rod to maximize the probability of a first Compton interaction. A high-Z Cadmium Zinc Telluride (CZT) assembly is used to absorb the scattered X-rays in photoelectric effect interactions (Fig. 2). Owing to the properties of the Klein-Nishina scattering cross section, the azimuthal scattering angle distribution peaks at angles +/- 90° from the preferred orientation of the electric field vector.

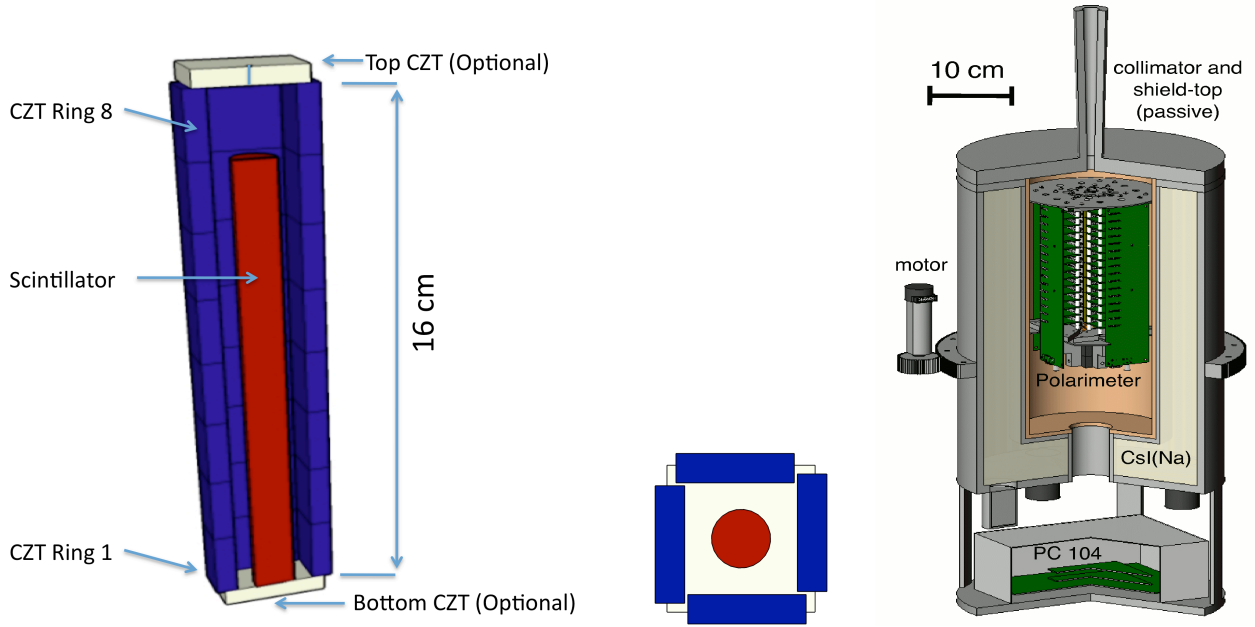

Figure 2: Left and center panels: conceptual design of the X-Calibur detector assembly consisting of a scattering slab (red) surrounded by absorbing CZT detectors (blue and white). The scattering slab is aligned with the optical axis of an X-ray mirror (not shown here). A large fraction of the Compton scattered photons are subsequently photo-absorbed in the high-Z CZT detectors. The distribution of the azimuthal scattering angles constrains the linear polarization of the incoming radiation. In the left panel, the white rectangles at the top (front-side, with a cylindrical bore in the center) and bottom (rear end) of the detector assembly show optional additional CZT detectors discussed in Sect. 5.3. The right panel shows the polarimeter with the readout electronics inside a fully active CsI shield which can be rotated around the optical axis.

The low-Z/high-Z combination leads to a high fraction of unambiguously detected Compton events and suppresses the detection of elastically scattered neutrons that can mimic Compton events. The first X-Calibur flight will use a scintillator as the scattering rod. The scintillator is triggered when sufficient energy is deposited and read out by a PMT, allowing us to look for coincidences in the scintillator and CZT detectors and to get a better understanding of the background level. We will use the data gathered during the flight to perform detailed comparisons between the simulated and measured energy spectra in the CZT detectors for ON (source + background) and OFF (background only) observations. Simulated and observed background data will test the predictive power of our background model by plotting the simulated and measured CZT energy spectra: (i) for all events detected in the CZT detectors, (ii) for all events detected in the CZT detectors with and without a trigger in the scintillator slab, (iii) for all events detected in the CZT detectors with and without a shield veto, and (iv) for all events detected in the CZT detectors with and without a trigger in the scintillator slab, and for both event classes with and without a shield veto.

The length of the scintillator of X-Calibur is 14 cm. It was chosen to yield a Compton scattering probability of >90% for 75 keV photons. We performed a dedicated Monte Carlo study to optimize the diameter $d$ of the scattering slab, accounting for the point spread function of the X-ray mirror, the defocus of the X-ray beam away from the focal point, and alignment errors (see Table 1). Figure 3 shows the Figure of Merit (FoM) as function of $d$. The FoM is chosen to be proportional to the modulation factor and the square root of the detection efficiency (the fraction of incident photons being scattered and reaching the CZT detectors). In the signal-dominated regime, this FoM will be inversely proportional to the MDP. A thicker scattering rod catches a higher fraction of photons and thus leads to a higher detection efficiency. We infer the azimuthal scattering angle from the position of the triggered CZT pixel assuming that the X-ray scattered at the optical axis. For a thicker scattering rod the spatial uncertainty of the scattering location leads to larger errors in the inferred azimuthal scattering angle. As a consequence, the modulation factor decreases. Figure 3 shows that the FoM exhibits a broad maximum for diameters between 1 cm and 1.3 cm. For our simulations described below, we assume a diameter of 1 cm.

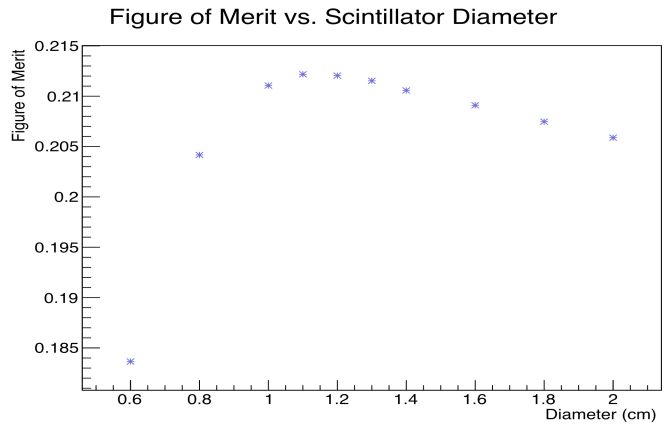

Figure 3: Figure of Merit values plotted vs. Scintillator Diameter. The peak is at diameter around 1.1 cm.

Table 1: Budget of errors important for the optimization of the thickness of the scattering slab for a one-day balloon flight of X-Calibur on InFOCµS.

| Error Source | Error [arcmin] | Error [mm] |
|---|---|---|
| PSF 75% Cont. Radius | 2 | 4.6 |
| Max. Defocus | <0.85 | <2 |
| Max. Pointing Error | <1 | <2.3 |
| Max. Align. Error | <0.85 | <2 |
| Total | <4.9 | <10.6 |

The CZT detector configuration is made of 32 detector units (each $0.2\times2\times2$ cm$^3$, more units for longer scattering rods) with a monolithic cathode oriented towards the inside of the assembly and $8\times8$ anode pixels each (2.5 mm pitch) oriented towards the outside. The length of the CZT assembly is longer than the scattering rod to catch a high fraction of the scattered X-rays. In the following we assume that the CZT detectors will operate at an energy threshold of 15 keV. Such a threshold is sufficiently low for a balloon flight, as <20 keV X-rays are almost completely absorbed in the atmosphere. For a satellite borne experiment, one would like to go to a lower energy threshold of about ~1-2 keV. As mentioned above, we infer the azimuthal scattering angle from the position of the CZT pixel with the highest signal assuming that all photons scatter at the optical axis. The residual atmosphere at balloon altitudes (2.9 g/cm$^2$) absorbs X-rays below 20 keV and the X-Calibur mirror is limited to $\leq 75$ keV energies. Below we will thus plot our results only in the 20-75 keV energy range. The degree of linear polarization can be measured based on events with (or without) a trigger of the PMT reading out the scintillator rod. The events with a scintillator hit have a somewhat lower level of background contamination. In X-Calibur, there is no crosstalk between the low-Z Compton scatterer and the high-Z photoelectric-effect absorber as these two detector elements are well separated from each other. Some of the alternative Compton polarimeter designs suffer from optical and/or electronic crosstalk that can lead to the misclassification of background events as Compton-events.

The polarimeter and the front-end readout electronics will be located inside an active CsI (Na) anti-coincidence shield with a passive top (Fig. 2) to suppress charged and neutral particle backgrounds. In the simulations, we assume the active shield is 5 cm thick and the passive shield is 2 cm thick. The active shield is read out by 4 PMTs with a high quantum efficiency super-bi-alkali photo cathodes placed at the rear end of the shield. The PMT trigger information allows to effectively reject background from signal events.

## 4  Simulations

This section describes the simulations for the optimization of the X-Calibur design. For this purpose, we simulated polarized and unpolarized X-ray beams and the most important background sources. The simulations account for the atmospheric absorption at float altitude (130,000 feet) and for the mirror throughput.

### 4.1  Simulation Details

Our Monte Carlo study uses the GEANT4 simulation package [22] and a detector simulation code. We used the GEANT4 package with the Livermore Low-Energy Electromagnetic Models [23] to simulate 2 million polarized and 2 million unpolarized photons. Photons with energies between 20 keV and 75 keV were generated according to the Crab spectrum as measured with the Swift Burst Alert Telescope (BAT) telescope [24]:

$$\frac{dN}{dE} = 10.17 \left(\frac{E}{1\ keV}\right)^{-2.15} \text{ph cm}^{-2}\ s^{-1}\ keV^{-1}. \tag{3}$$

We account for atmospheric absorption at a floating altitude of 130, 000 feet using the NIST XCOM attenuation coefficients for an atmospheric depth of 2.9 g/cm$^2$ [25]. The simulations assume observations at a zenith angle of $10°$. The left panel of Fig. 4 shows the atmospheric transmissivity. It increases rapidly from 0 to 0.6 in the 20 - 60 keV range. The right panel of Fig. 4 shows the effective area of the Nagoya University Al mirror as function of the photon energy. The throughput is 95 cm$^2$, 60 cm$^2$, and 40 cm$^2$ at 20 keV, 30 keV, and 40 keV, respectively.

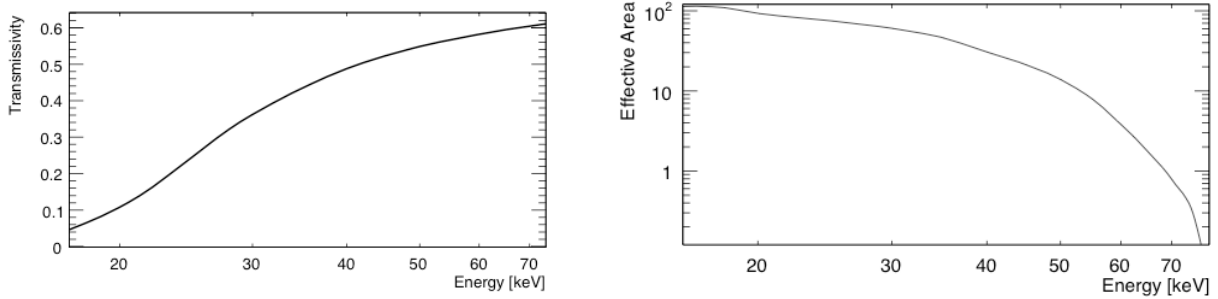

Figure 4: The X-ray transmissivity at a floating altitude of 130,000 feet calculated with the NIST XCOM attenuation coefficients [25] for an atmospheric depth of 2.9 g/cm$^2$ (left panel). The effective area as function of the photon energy of the mirror that will be used in the first InFOCμS balloon flight planned for 2013 (right panel).

The incident photons and their secondaries are tracked through the detector volumes recording interaction locations and energy depositions occurring along the way. The detector simulation code determines the energy deposited in the pixels of individual detectors, and generates a signal if the energy exceeds the trigger threshold. If an event triggers more than one detector element, all the energy detected in adjacent detector elements is summed together with the highest energy deposited. We use the position information of highest energy deposited in the analysis.

## 4.2  Background simulations

In this section, we report on background simulations. We simulated the most important backgrounds such as the Cosmic X-ray Background (CXB) [26], secondary gamma rays (upward and downward components), cosmic ray protons and electrons (of primary and secondary origin, the primaries move downwards, the secondaries move upward and downward) [27] with the MEGALIB software package [28]. The neutron background was not modeled since the detailed studies of Parsons et al. [29] showed that neutrons do not contribute substantially to the background in CZT detectors [29]. The models cover the entire solid angle (4π sr) and the energy range between 20 keV and 100 GeV. In our background simulations, the incident particles were generated on a spherical surface of 46 cm radius surrounding the detector model.

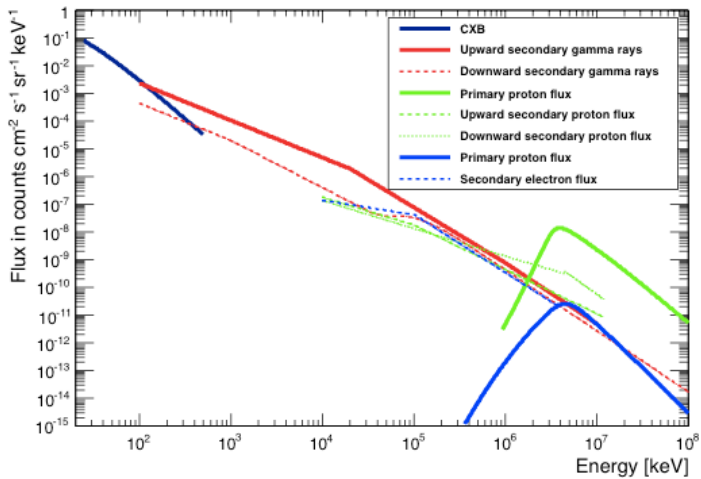

Figure. 5 This figure shows the energy spectra of the simulated background sources. The CXB (black curve) is from M. Ajello et al. [26], the red solid curve represents upward secondary gamma rays [27]. The red dashed curve represents downward secondary gamma rays [27]. The green solid curve is the primary proton flux [27], and the green dashed curve is the secondary proton flux [27]. The blue solid line is the primary electron flux [27] and the blue dashed line is the secondary electron flux [27].

Fig. 5 shows the background input spectra used in the simulations. Below a few hundred keV, the incident background is dominated by the CXB for all directions not shielded by the Earth. From 200 keV to ~400 MeV, the upward-moving secondary gamma rays resulting from cosmic ray interactions in the atmosphere dominate the background. Above 10 GeV, cosmic ray (CR) protons are the main contributors to the background. Different shield configurations and shield thicknesses were simulated to optimize the shield design. We compared the background level for different shield designs, i.e. fully active shields with wall thicknesses between 2 cm and 10 cm and active/passive shield combinations.

The configuration shown in the right panel of Fig. 2 represents a compromise, balancing the background rejection power with the mass and complexity of the shield. The shield combines a 5 cm active portion with a 2 cm thick passive Pb-cover, and a Pb-collimator.

## 5 Expected Performance

### 5.1 Azimuthal Scattering Distributions

The left panel of Fig. 6 shows exemplary azimuthal scattering distributions for unpolarized and polarized incident X-ray beams before correction for non-uniformities. The results correspond to 5.6 hours of on-source observations of the Crab pulsar and nebula during a one-day balloon flight of the experiment from Fort Sumner using the Nagoya University Al mirror. Even for an unpolarized incident X-ray beam, the φ-distribution shows a modulation owing to the large pixel size (2.5 mm) and associated aliasing effects. Before computing the MDP with Eqs. (1) and (2) we correct for binning effects by dividing the polarized distributions by the unpolarized distributions. The correction flattens the φ-distributions of the unpolarized beams and leads to a sinusoidal modulation of the φ-distributions of the polarized beams (see right panel of Fig. 6). See [1] for a detailed description of the correction procedure and for a study of the validity of Eq. (2) when the correction is used. X-Calibur achieves modulation factors of ~0.5 (see Table 3, and Fig. 8, left panel) typical for Compton polarimeters.

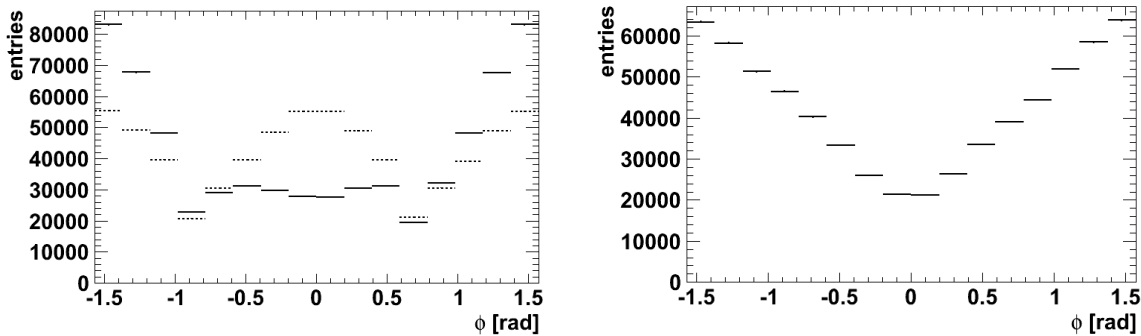

Figure. 6. The left panel shows distribution of azimuthal scattering angles for a polarized beam (solid lines) and an unpolarized beam (dashed lines). The right panel shows distributions of azimuthal scattering angles for a polarized X-ray beam after correcting for binning effects. All events triggering one or more CZT detectors have been used in the analysis.

## 5.2 Performance with 4 different scattering materials

We compared the performance achieved with a scintillator scattering rod with that achieved with scattering rods made of alternative materials, i.e. Be, Li, and LiH. Table 2 lists the physical characteristics and dimensions of the different scattering rods and the CZT detector assemblies for each case. When a plastic scintillator is used as the scattering material, the rod is read out with a photomultiplier. The other three scattering materials are assumed to be passive. The length of all scattering rods was chosen to yield a Compton scattering probability of 90% for 80 keV photons. The most important results, including the rates of Compton events for a Crab like source $R_{Crab}$ [Hz], the peak detection efficiency and the energy at which this efficiency is achieved, the modulation factor μ and the Minimum Detectable Polarization MDP, are discussed below and are summarized in Table 3.

Note that the background for most competing polarimeter designs is indeed high as they have wide field of views and use very massive detectors. However, the X-Calibur background is much lower – as we use focusing optics and a thick shield. The residual X-Calibur background is almost negligible for strong sources (like the Crab Nebula). For such sources (i.e. all the sources that X-Calibur will observe during the first balloon flight), it pays off to use all events – even if they do not have a trigger signal in the scintillator. For weak sources and long observations it becomes advantageous to use the scintillator trigger. For example, for a $10^6$ s observations of a 25 mCrab source, the MDP is 4.5% when we require the scintillator to be triggered while it is 5.5% for the case without scintillator trigger requirement.

Table 2. Sizes of different scattering rods and CZT detector (0.2×2×2 cm$^3$ each) assemblies discussed in the text.

|  | Design 1 | Design 2 | Design 3 | Design 4 |
|---|---|---|---|---|
| Scatterer | Scintillator EJ-200 | Be | LiH | Li |
| Diameter | 1 cm | 1 cm | 1 cm | 1 cm |
| Length of Scintillator | 14 cm | 9 cm | 18 cm | 32 cm |
| CZT Unit | 0.2×2×2 cm$^3$ | 0.2×2×2 cm$^3$ | 0.2×2×2 cm$^3$ | 0.2×2×2 cm$^3$ |
| Length of CZT assembly | 16 cm | 12 cm | 20 cm | 34 cm |

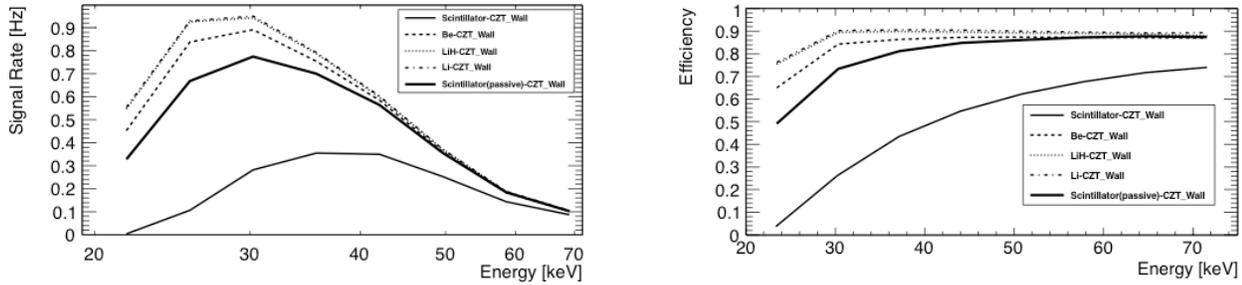

Figure 7: The left panel shows the detection rate of Compton events for scattering rods made of various materials (assuming a balloon flight with X-Calibur and the Nagoya University mirror at an altitude of 130,000 feet). The calculation assumes a source with a Crab-like flux and energy spectrum. The different lines show the results for the different scattering rods and, for the scintillator rod, for different trigger requirements. The right panel presents the energy dependence of the detection efficiency (number of detected events divided by the number of photons incident on the polarimeter).

Figure 7 shows the detection rates and the detection efficiencies achieved with the different scattering rods. In the left panel, the thick solid line shows the detection rates without the trigger requirement in the scintillator (i.e. using the rod as a passive scatterer) while the thin solid line shows the detection rates with a 2 keV threshold in the scintillator. The efficiency is defined here as the fraction of the photons impinging on a detector assembly that trigger the instrument and enter the polarization analysis. The simulations show that the lower-Z materials lead to higher rates and efficiencies, especially at <50 keV energies. The scintillator (solid line) gives the lowest rates and efficiencies while LiH (dash-dotted line) gives the highest. The high efficiencies of the X-Calibur design close to 100% can be explained by the fact that all source photons hit the scattering rod, a large fraction of the photons Compton scatter in the low-Z material, and most of the scattered photons can escape the rod and are detected in the CZT detectors. The modulation factors are close to 0.5 for all four scattering rods (Fig. 8, left panel). The LiH scatterer achieves the best MDP (3.02%) followed by Li (3.06%), Be (3.13%), and the Scintillator (4.5% when scintillator triggered at a 2 keV threshold while 3.41% when using it as passive scatterer) (Fig. 8, right panel). The lower-Z materials perform better, but the scintillator can be read by a PMT to give a coincidence signal to identify proper Compton events. As mentioned above, we use the scintillator for the first X-Calibur flight as it allows us to perform detailed tests of the performance and backgrounds of the polarimeter. Note that the <25 keV polarimeter response is relatively more important in space than for a balloon-borne mission (in the latter case, the atmosphere absorbs most of the <25 keV flux). In a space borne mission, a LiH scattering rod performs much better than a scintillator rod and would be our preferred choice.

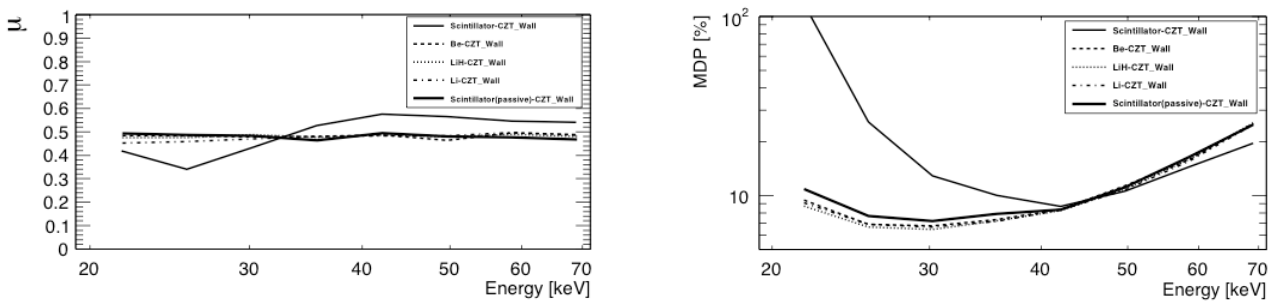

Figure. 8. Comparison of the modulation factor (left panel) and the MDP (right panel) for a one-day 5.6 hrs on-source observation of the Crab Nebula with X-Calibur and the Nagoya University mirror. The binning was chosen to have 8 statistically independent bins spaced logarithmically between 20 keV and 75 keV. The different lines show the results for different scattering materials. The lines between the data points are only shown to guide the eye.

Table 3. Comparison of the performance achieved with the four different scattering materials.

|  | Scintillator-CZT | Scintillator (passive)-CZT | Be-CZT | LiH-CZT | Li-CZT |
|---|---|---|---|---|---|
| $R_{Crab\ (InFOC\mu S)}$ [Hz] | 1.58 | 3.68 | 4.17 | 4.46 | 4.49 |
| Peak efficiency | 0.76 (69 keV) | 0.87 (65 keV) | 0.86 (70 keV) | 0.87 (65 keV) | 0.88 (70keV) |
| μ | 0.52 | 0.48 | 0.48 | 0.48 | 0.47 |
| MDP [a] | 4.50 % | 3.41% | 3.13 % | 3.06% | 3.02 % |

[a] The MDP is given for a 5.6 hrs balloon flight of X-Calibur and InFOCμS with the 40 cm diameter mirror from the Nagoya University .

## 5.3 Performance with additional CZT detectors

In this section we discuss the performance improvement achieved with additional CZT detectors placed at the two ends of the CZT detector assembly (see the white rectangles in the left panel of Fig. 2). We use the same assumptions as in the previous sections. Fig. 9 shows the simulated detection rates: (i) in the standard detector assembly, (ii) in the additional CZT detector at the front-side of the detector assembly, (iii) in the additional CZT detectors at the rear end. The simulation results show that the detection rates in the additional CZT detectors are very low. The reference X-Calibur design thus does not include the additional detectors.

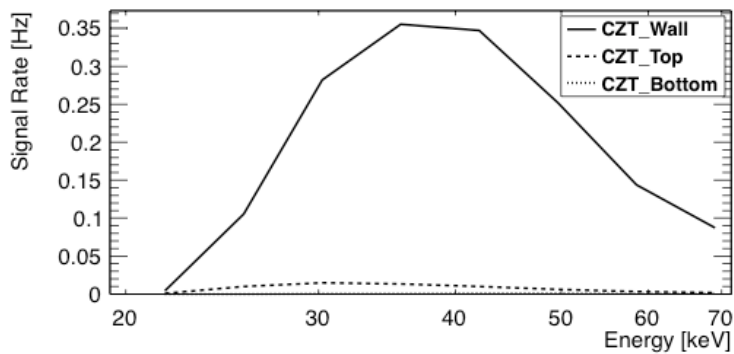

Figure. 9. Comparison of detection rates in the CZT walls and in the additional CZT detectors at the front-side and rear end of the CZT detector assembly of the reference design (Fig. 2).

## 5.4 Performance of the 8 CZT detector rings

The CZT detector configuration of X-Calibur is made of 8 rings of four detectors each (32 detector units altogether). In the following, we number the detector rings from 1 at the rear end to 8 at the front end of the assembly (Fig.2). The left panel of Fig. 10 shows the detection rates for all 8 rings. The right panel of the same figure shows the MDPs achieved with the different rings. Owing to the geometry of the scintillator/CZT detector configuration, ring #7 sees most events as it detects forward, backward, and sideward scattered photons. From ring #6 to ring #1, fewer and fewer events are detected, as most of the events are scattered near the front end of the scintillator. Ring # 8 sees a substantial number of events, but less than ring #7, because it can only detect backscattered photons.

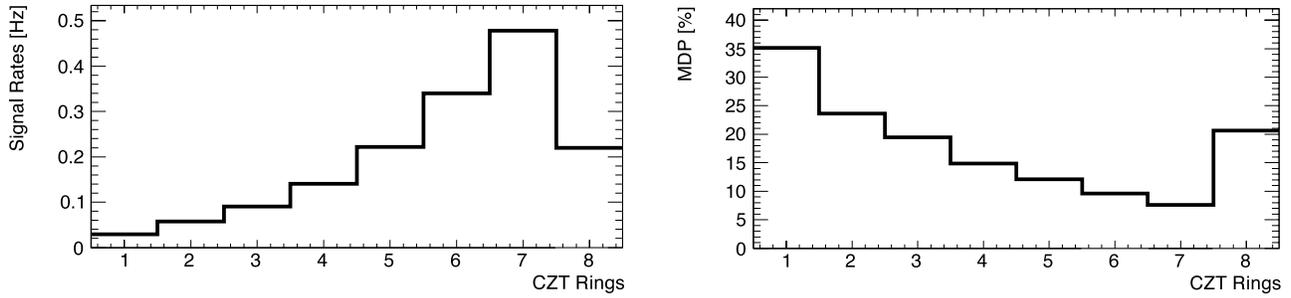

Figure. 10. Comparison of detection rates (left panel) and MDPs (right panel) in different CZT Rings. Ring #1 is at the rear end of the detector assembly and ring #8 at the front-side

We simulated the energy spectra detected in each ring of CZT detectors to study how the energy resolution deteriorates owing to the fact that the primary photons deposit some energy in the scintillator, which we cannot measure accurately owing to the scintillator's poor energy resolution. Photons with energies of 20, 60, and 80 keV were generated and the simulations accounted for an inherent RMS energy resolution of the CZT detectors of 2 keV. Fig. 11 shows energy spectra for 20 and 60 keV photons, and Fig. 12 shows the energy resolution for each detector ring. The energy resolution is better at lower photon energies owing to the smaller relative fraction of energy lost to the Compton electron. The CZT ring #8 has a better energy resolution than the other 7 CZT rings because it only sees backscattered photons with a narrow range in scattering angles.

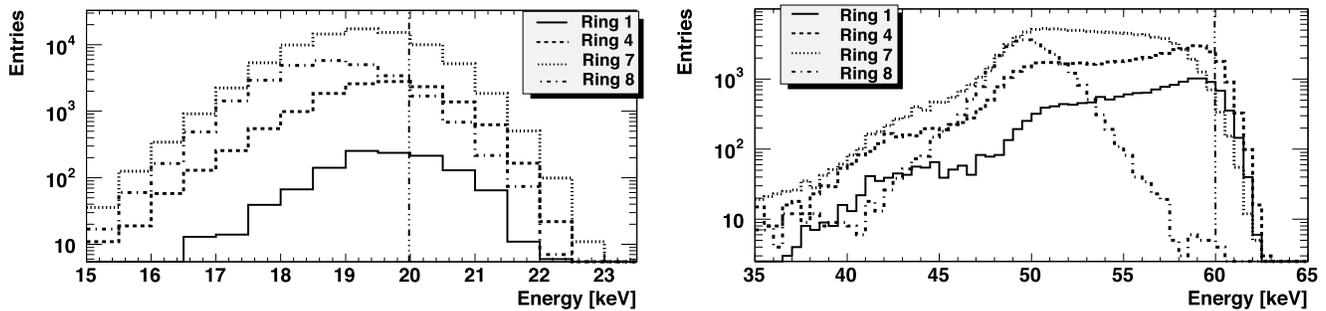

Figure. 11. The left panel shows the energy depositions measured in CZT rings # 1, # 4, # 7 and # 8 when a 20 keV photon beam Compton scatters in the scintillator. The right panel shows the same for a 60 keV photon beam. The simulations include an energy resolution with a RMS of 2 keV.

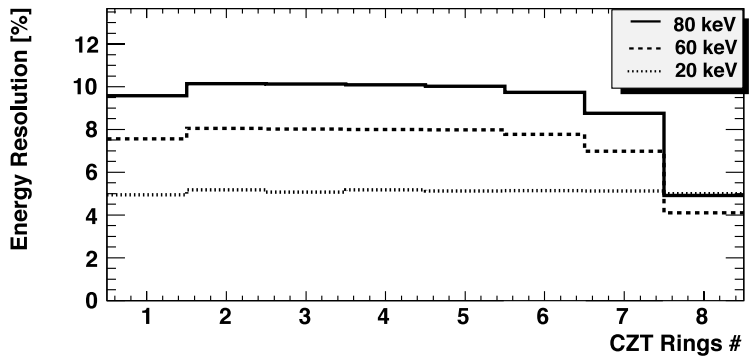

Figure. 12. Energy resolutions of each CZT ring for 20 keV, 60 keV and 80 keV photons. The energy resolution is the RMS value of the detected energy divided by the mean amplitude.

## *5.5* Optimization of the Shield Design

We computed the residual background level for different shield designs. We compared the performance achieved with the 5 cm active CsI shield with that of a passive Pb shield with the same mass (wall thickness 1.96 cm). The left panel of Fig. 13 shows the background rates in the CZT detectors for the two shields. The active shield used with a >50 keV active veto outperforms the passive shield of the same mass by one order of magnitude in terms of the residual background rate. We simulated active shields with wall thicknesses of 3, 5, and 7 cm. The right panel of Fig. 13 shows the residual background energy spectrum for these shield configurations. The residual background rate decreases monotonically with the thickness of the active shield. For our balloon flight, the dominating background comes from the albedo photons (upward secondary gamma rays) from the earth side. Our simulation studies show that a thick active shield suppresses this background most effectively. The balloon flight results from Slavis et al. [30, 31] back up the results. Balancing the background rejection power with the mass and complexity of the shield, we choose a 5 cm CsI (Na) shields with a 2 cm passive front shield as our reference design (Fig. 2).

Table 4 shows the contributions of different background components to the total background rate for the reference design and a veto energy threshold of 50 keV. As expected, the secondary gamma rays are the main contributor to the residual background rate. Fig. 14 shows the residual background rate as function of the CsI (Na) energy threshold. The rate increases with the veto threshold of the shield. We choose 50 keV as the veto trigger threshold of the CsI (Na) active shield leading to a residual background rate of 0.007 Hz. The veto rates in the active shields are 57.21 Hz. Assuming that each background hit vetoes the CZT detectors for 5 μsec, the background will cause a dead time of 0.03%.

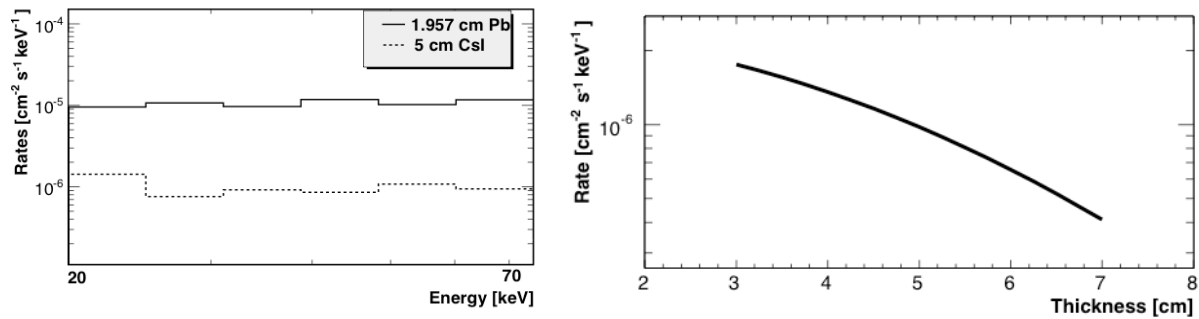

Figure 13. Left panel: Residual background rate for the active/passive shield design. Right panel: Residual background rate of the "reference design" ("active CsI shield, passive Pb front end") as function of the thickness of the active shield.

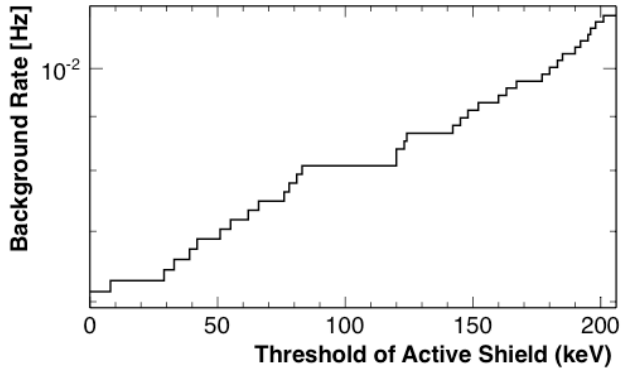

Figure. 14. The residual background rate as function of the CsI (Na) veto threshold for the same flight altitude. We assume a veto threshold of 50 keV for our sensitivity estimates.

## *5.6* Predicted Sensitivity of X-Calibur when Flown on Balloons

Table 4 The contributions of different background components to the total background rate for the reference design and a veto energy threshold of 50 keV

| Background contributor | CXB [$cm^{-2}s^{-1}keV^{-1}$] | Secondary gamma rays [$cm^{-2}s^{-1}keV^{-1}$] | Protons [$cm^{-2}s^{-1}keV^{-1}$] | Electrons [$cm^{-2}s^{-1}keV^{-1}$] |
|---|---|---|---|---|
| Rates (with scintillator trigger) | $1.3 \times 10^{-8}$ | $9.8 \times 10^{-7}$ | $6.5 \times 10^{-11}$ | $2.1 \times 10^{-9}$ |
| Rates (no scintillator trigger) | $6.8 \times 10^{-8}$ | $2.0 \times 10^{-5}$ | $3.9 \times 10^{-10}$ | $2.8 \times 10^{-8}$ |

On the one-day balloon flight from Fort Sumner we expect to be able to observe several sources, including the Crab pulsar and nebula, for a few hours. Simulations performed for sources at different zenith angels θ, show that $R_{src}$ scales proportional to $(\cos\theta)^{1.3}$. We use this scaling law when estimating MDPs. Assuming a 5.6 hrs observation of the Crab X-Calibur achieves a Minimum Detectable Polarization degree (MDP, 99% confidence level) of 4.5%. Fig. 15 shows the outcome of a 5.6 hrs on-source observation of the Crab Nebula (from top to bottom: Reconstructed flux, reconstructed polarization degree, and reconstructed polarization direction) with X-Calibur and the Nagoya University mirror for a balloon flight altitude of 130,000 feet. The graphs show that the observations will give precise measurements of the polarization degree and the polarization direction in several independent energy bins. We plan to follow up on the first balloon flight with proposals for Long Duration Balloon (LDB) flights. An attractive option is a LDB flight with a larger X-ray mirror assembly and then X-Calibur would extend the spectropolarimetric coverage to >60 keV. For a 3-day balloon flight from Alice Springs, X-Calibur/InFOCµS achieves a Minimum Detectable Polarization degree (MDP, 99% confidence level) of 2.60% for a 16.8 hrs observation of a source with a Crab-like flux and energy spectrum. For a 3-week balloon flight from McMurdo, estimating the background to be about 5.5 times higher than that of the balloon flight from Fort Sumner, the MDP would go down to ~1.0% in a 117 hrs on-source observation. Additional mirror modules could reduce the MDP well below the 1% limit.

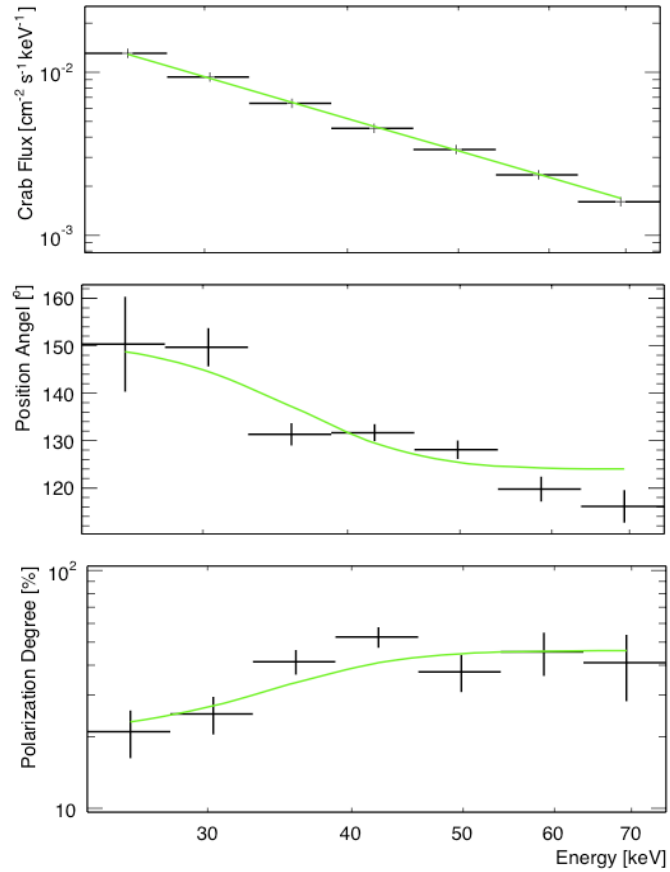

Fig. 15. Simulated outcome of a 5.6 hrs on-source observation of the Crab Nebula (from top to bottom: Reconstructed flux, reconstructed polarization degree, and reconstructed polarization direction) with X-Calibur and the Nagoya University mirror for a balloon flight altitude of 130,000 feet. The green lines show the assumed model distributions: the Crab spectrum is from [24]; we assume the polarization degree and direction change continuously from the values measured at 5.2 keV with OSO-8 [21] to the values measured at >100 keV with INTEGRAL [32].

## 6  Summary and Discussion

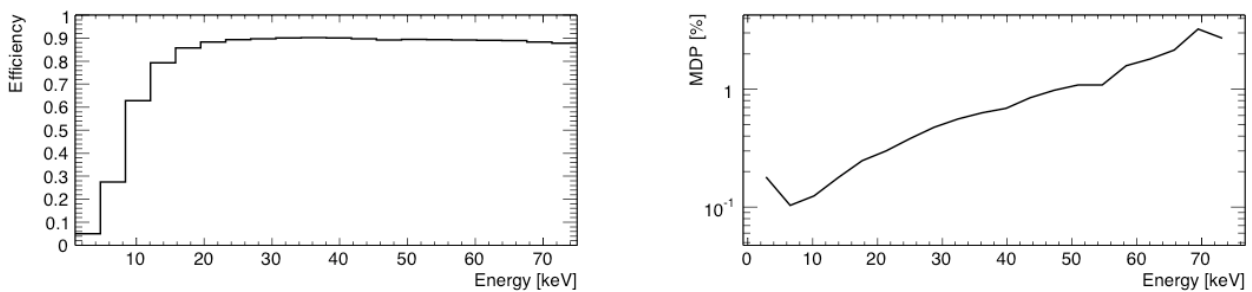

,
Figure 16: Energy dependence of the detection efficiency (number of detected events divided by the number of photons incident on the polarimeter) and MDP of Compton events for a satellite-borne version X-Calibur with LiH scattering rod and CZT detectors with a 1 keV energy threshold for a $10^6$ s observations of a 100 mCrab source with a Crab spectrum using the background estimates derived for the NuSTAR mission [33]. The binning was chosen to have 20 statistically independent bins spaced logarithmically between 1 keV and 75 keV. The MDP calculation assumes 6 identical polarimeters, each using a NuSTAR mirror.

In this paper we describe the optimization of the design of the X-Calibur Compton polarimeter by comparing the performance achieved with four different Compton scattering materials (Scintillator, Be, LiH, Li), with different CZT detector assemblies, and with several shield configurations.

The conclusions from our study can be summarized as follows:

- X-Calibur combines a detection efficiency of order unity with a high modulation factor of $\mu\sim0.5$. The detection efficiency and the modulation factor have values close to the maximum theoretically possible values given the physics of Compton scatterings.
- The lower-Z scattering rods (i.e. (Be, LiH and Li) perform better than higher-Z scattering rods (e.g. made of scintillator), especially at <20 keV energies. As the atmosphere absorbs most of the <20 keV X-rays, we decided to use a scintillator rod for the first balloon flight. The trigger information from the scintillator will allow us to perform additional tests of our simulation model.
- Additional CZT detectors at the front and rear ends give only a marginal improvement of the detection efficiency.
- On a balloon-borne mission, upward secondary gamma rays dominate the non-vetoed background rate. The simulations predict a non-vetoed background rate of $9.8\times10^{-7}\,\mathrm{cm}^{-2}\,\mathrm{s}^{-1}\,\mathrm{keV}^{-1}$ (with a >2 keV energy deposition in the scintillator rod and a 20~75 keV energy deposition in a CZT detector) and $2.0\times10^{-5}\,\mathrm{cm}^{-2}\,\mathrm{s}^{-1}\,\mathrm{keV}^{-1}$ (a 20~75 keV energy deposition in a CZT detector while no scintillator trigger requirement).
- We derive excellent Minimum Detectable Polarization degrees (MDP, 99% confidence level) of the Crab Nebula: 3.41% for a 1-day balloon flight from Fort Sumner with 5.6 hrs ON-source observation; 2% for a 3-day balloon flight from Alice Springs with 16.8 hrs ON-source observation; 0.7% for a 3-week balloon flight from McMurdo with 117.6 hrs ON-source observation.

Successful balloon flights would motivate a satellite-borne mission. For a space-borne mission, a low-Z LiH scattering rod would make it possible to achieve excellent sensitivity over the broad energy range from ~2 -70 keV. We estimated the detection efficiency and the MDP for a satellite-borne mission with 6 NuSTAR-type mirrors and 6 identical X-Calibur-type polarimeters (see [33] for a similar mission concept). The calculations were performed for a $10^6$ s observations of a 100 mCrab source assuming CZT detectors with a 1 keV energy threshold. We used the results of the simulations carried through for NuSTAR to estimate the background level [34]. Fig.16 shows that such a Compton polarimeter would have appreciable sensitivity over the wide energy range from ~1 keV to 70 keV.

# 7 Acknowledgments

We acknowledge NASA for support from the APRA program under the grant NNX10AJ56G and support from the high-energy physics division of the DOE. The Washington University group is grateful for discretionary funding of the X-Calibur polarimeter by the McDonnell Center for the Space Sciences. Q.G. thanks the Chinese Scholarship Council from China for the financial support (NO.2009629064) during her stay (from August 2009 to July 2011) at Washington University in St Louis.